%% file: main.tex
\documentclass[sigconf, 9pt]{acmart}

\AtBeginDocument{%
	\providecommand\BibTeX{{%
			\normalfont B\kern-0.5em{\scshape i\kern-0.25em b}\kern-0.8em\TeX}}}

\renewcommand\footnotetextcopyrightpermission[1]{} 

\usepackage{xspace}
\newcommand{\Design}{HyperOMS\xspace}
\newcommand{\mz}{$m$/$z$\xspace}

\usepackage{mathtools,xparse}

\usepackage{enumitem}
\setlist[itemize]{noitemsep, topsep=0pt, leftmargin=*}
\usepackage{cprotect}
\usepackage{subcaption}
\usepackage{epstopdf}
\usepackage{fancyvrb}

\usepackage{multirow}
\usepackage{multicol}
\usepackage{dsfont}

\newcommand{\blackcircle}{\raisebox{-.6ex}{\scalebox{2.30}{$\bullet$}}}
\def\invcircledast#1{%
	\mathbin{\vphantom{\circledast}\text{%
			\ooalign{\smash{\blackcircle}\cr
				\hidewidth\smash{\textcolor{white}{\bf \footnotesize $#1$}}\hidewidth\cr
			}%
	}}%
}

\newcommand{\squeezeup}{\vspace{-\baselineskip}}
\newcommand{\smallsqueezeup}{\vspace{-.8\baselineskip}}
\begin{document}
\setlength{\belowdisplayskip}{0pt} \setlength{\belowdisplayshortskip}{0pt}
\setlength{\abovedisplayskip}{0pt} \setlength{\abovedisplayshortskip}{0pt}

\title{Massively Parallel Open Modification Spectral Library Searching with Hyperdimensional Computing }

\author{Jaeyoung Kang, Weihong Xu, Wout Bittremieux, Tajana Rosing}
\affiliation{%
	\vspace{0.0em}\institution{University of California, San Diego}\country{}
}

\email{%
	{j5kang, wexu, wbittremieux, tajana}@ucsd.edu
}

\begin{abstract} 
Mass spectrometry, commonly used for protein identification, generates a massive number of spectra that need to be matched against a large database. In reality, most of them remain unidentified or mismatched due to unexpected post-translational modifications. Open modification search (OMS) has been proposed as a strategy to improve the identification rate by considering every possible change in spectra, but it expands the search space exponentially. In this work, we propose \Design, which redesigns OMS based on hyperdimensional computing to cope with such challenges.
Unlike existing algorithms that represent spectral data with floating point numbers, \Design encodes them with \textit{high dimensional binary vectors} and performs the efficient OMS in high-dimensional space. With the massive parallelism and simple boolean operations, \Design can be efficiently handled on parallel computing platforms. Experimental results show that \Design on GPU is up to $17\times$ faster and $6.4\times$ more energy efficient than the state-of-the-art GPU-based OMS tool~\cite{annsologpu} while providing comparable search quality to competing search tools. 
\end{abstract}
\input{./ccs}
	

\maketitle
\pagestyle{plain}

\smallsqueezeup
\section{Introduction}\label{sec:introduction}
\begin{figure}[t]
	\centering
	\includegraphics[width=.8\linewidth]{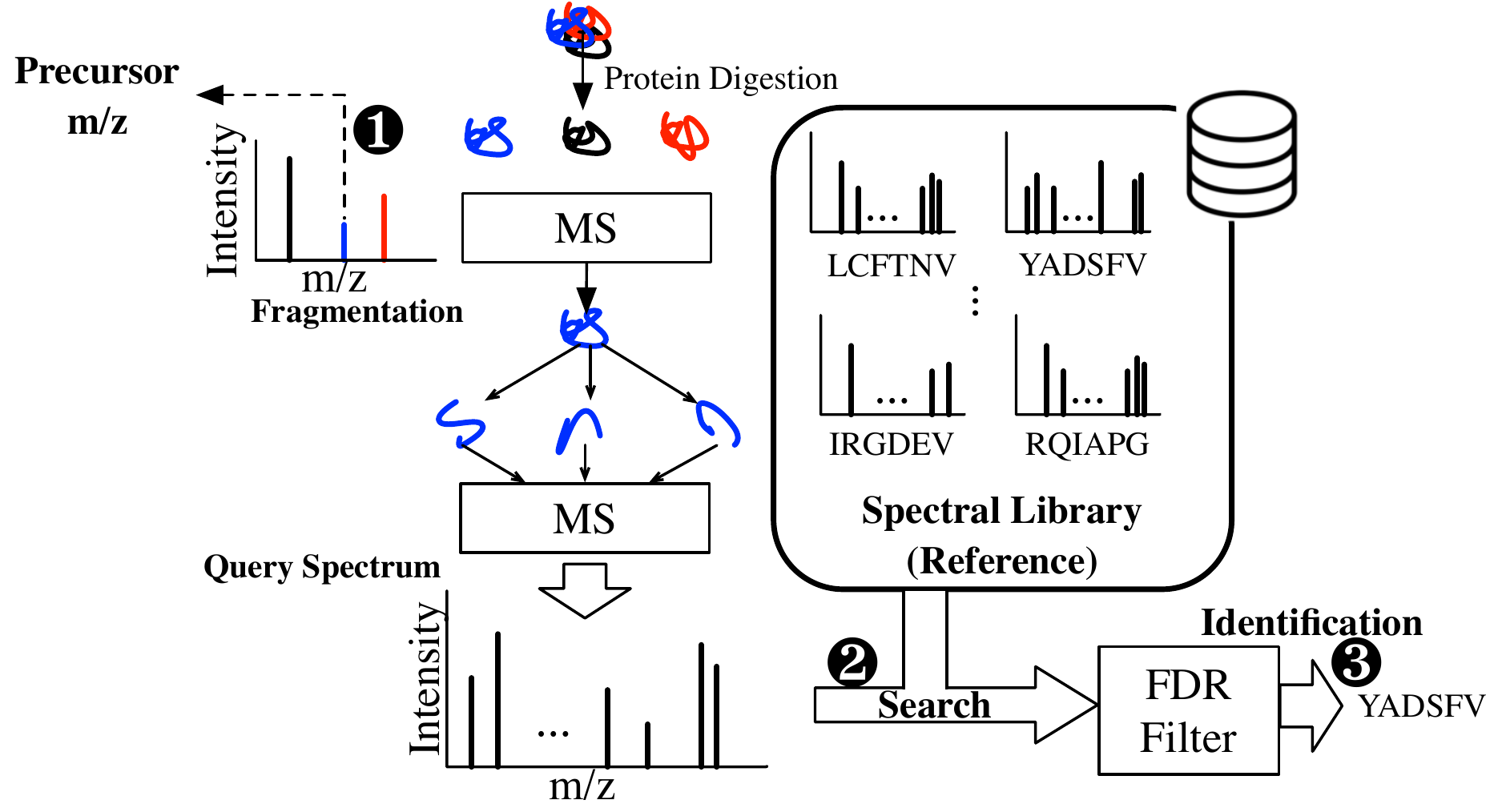}
	\caption{Overview of spectral library searching. Standard searching uses a narrow precursor \mz tolerance, while OMS uses a wide precursor \mz tolerance during the searching. } 
	\squeezeup
	\label{fig:spectrum}
\end{figure}

Proteomics plays an essential role in understanding the molecular mechanisms of proteins, which are responsible for various tasks in a life of a cell~\cite{Aebersold2016}.
Mass spectrometry (MS) is one of the most popular and reliable approaches to identifying and quantifying proteins and peptides in complex biological samples~\cite{Smith2014}. 
In a typical MS experiment, a method called tandem mass spectrometry (MS/MS) generates a massive amount of MS/MS spectra data. 
Then researchers determine peptide annotations of the MS/MS spectra via spectral library searching~\cite{Lam2011}. 
Peptide sequences are assigned to experimental MS/MS spectra by matching them against a spectral library of known peptides (see Figure \ref{fig:spectrum}).

The challenge of spectral library searching is that a significant portion of MS/MS spectra acquired during an experiment cannot be directly identified~\cite{Chick2015} using searching approaches with conventional similarities, such as cosine similarity. In reality, proteins undergo one or more post-translational modifications (PTMs), which change their mass and MS/MS fragmentation pattern. PTMs can be introduced during sample preparation as an artifact of MS measurement~\cite{Bittremieux2017}, or biologically relevant PTMs arise \emph{in vivo}~\cite{Walsh2005}. However, spectral libraries mainly contain reference spectra for unmodified peptides, so PTMs make experimental spectra challenging to identify as they no longer exactly match the reference spectra.

Open modification searching (OMS) is a promising approach to circumvent this limitation and identify modified spectra~\cite{Na2015}. 
Standard spectral library searching only compares experimental spectra to reference spectra with a similar precursor mass, i.e., the mass of the unfragmented peptide, as matching peptides should have an identical mass. 
In contrast, OMS performs spectra matching on a wider range of reference spectra, where OMS compares modified query spectra to their unmodified reference variants, even when their precursor mass differs due to PTMs.
OMS' wider searching range empowers higher identification capability, thus enabling the study of more complex protein interaction in virus-host~\cite{Adams2022} and proteomics analysis of non-model organisms~\cite{heck2020proteomics}.

OMS faces several challenges in terms of low searching speed, efficiency, and accuracy. Compared to standard searching, the OMS speed is extremely slow due to the drastically increased search space since it considers all possible reference matches for each query spectrum~\cite{bittremieux2018fast}. 
This problem is further exacerbated by the increasing spectral data due to the cost reduction in the MS experiment (2$\times$ in recent two years)~\cite{aksenov2017global, perez2022pride}. Also, large spectral libraries created by repository-scale mining of open MS data become available~\cite{Griss2013, wang2018assembling}. 
For example, the size of human HCD (higher
energy collisional dissociation) spectral libraries~\cite{massivekb, wang2018assembling} hold $2.15$ million data points, which is $4\times$ larger than the previous NIST-HCD~\cite{wang2018assembling}. 
MassIVE contains $5.6$ billion spectra, which corresponds to $448$TB in size~\cite{massive}.

Several tools have been introduced to efficiently perform OMS~\cite{msfragger, bittremieux2018fast, Chi2018, Devabhaktuni2019, spectrast, spectrastoms, annsologpu}. 
These tools use various techniques to refine the search space, such as fragment ion indexing~\cite{msfragger}, nearest neighbor searching~\cite{bittremieux2018fast, annsologpu}, or tag-based filtering~\cite{Chi2018, Devabhaktuni2019}.
For example, the state-of-the-art OMS tool \mbox{ANN-SoLo} performs nearest neighbor searching using GPU and computes shifted cosine similarities on candidates~\cite{annsologpu}. 
However, existing solutions involve a complex execution pipeline and exhibit low data parallelism requiring high-precision floating-point (FP32) arithmetic for good search quality, e.g., shifted cosine similarity~\cite{bittremieux2018fast}.
As such, we redesign an OMS algorithm that only involves hardware-friendly Boolean operations with a simple execution pipeline.

In this work, we propose novel hyperdimensional computing (HDC)-inspired OMS algorithm called \Design. The spectral matching algorithm of \Design is based on the efficient computing paradigm, HDC~\cite{Kanerva2009}, that has shown high efficiency for pattern matching tasks, especially on parallel computing platforms, e.g., GPU~\cite{kang2022xcelhd, kang2022openhd}.
HDC improves the data separability and robustness~\cite{poduval2021robust, zhang2021assessing} by mapping data into high-dimensional (HD) space, where the information is distributed to every dimension of the HD vector. We leverage HDC's robustness to minimize the effects of PTMs. 
In particular, our method reflects the spatial and value locality of peaks in the spectrum, making the encoded data resilient to peak shifts and intensity changes. 
\Design addresses the search space challenges in OMS by approximating possible MS peak changes; spectra can be identified with a single similarity computation.
Furthermore, the proposed HDC-empowered OMS algorithm is hardware-friendly as it replaces FP32 operations with simple Boolean arithmetic by encoding data into binary HVs, leading to better computational efficiency.
We implement the \Design algorithm on the high-performance HDC library \cite{kang2022xcelhd, kang2022openhd} on GPU, obtaining up to $17\times$ speedup and $6.4\times$ energy efficiency over the state-of-the-art OMS solution, ANN-SoLo~\cite{annsologpu} while offering comparable search quality.

\smallsqueezeup
\section{Background and Related Work}
\subsection{Mass Spectrometry Background}
\label{sec:background}

MS is used to study the biological process in proteomics via the analysis of protein expression or state in cells or tissue~\cite{Smith2014}. Proteins are ubiquitous building blocks of life, and they are composed of peptides, which are chains of amino acids, which can be described as a string of letters.

During MS data acquisition, peptides are ionized to receive a charge, and their mass-over-charge (\mz) is measured.
The first intact ions are measured in an MS scan using data-dependent acquisition, and the resulting MS spectrum contains the corresponding \mz values.
The most intense peaks in the MS spectrum are selected. It is further analyzed in MS/MS scans, i.e., the second mass spectrometer. Ions with matching \mz are isolated and fragmented to generate MS/MS spectra. Fragmentation occurs along the peptide backbone in between its constituent amino acids. Peptides are split into their possible amino acid subsequences.
We record the \mz and intensity values of all fragments, and the measured spectrum forms a \textit{unique fingerprint of the measured peptide}.
Thus, each MS/MS spectrum consists of fragment \mz values, \textit{spectrum charge}, and its intact \mz from the preceding MS scan, called the \textit{precursor \mz} (Figure~\ref{fig:spectrum}-$\invcircledast{1}$).

Spectral library searching determines which peptide corresponds to the measured spectra~\cite{Lam2011}. (Figure~\ref{fig:spectrum}-$\invcircledast{2}$). A spectral library contains reference spectra, each with known peptide labels. We first select the reference candidates with a similar precursor \mz to a query spectrum. Next, similarities between the query spectrum and all candidates are computed. Finally, the query spectrum is assigned the same peptide label as its highest-scoring reference match. Here, we apply a false discovery rate (FDR) filter on search results~\cite{Elias2007} (Figure~\ref{fig:spectrum}-$\invcircledast{3}$), which is a popular strategy called target--decoy strategy~\cite{Elias2007} in MS/MS analysis to reduce false positives.
Decoy spectra that cannot exist are added to the spectral library besides the real (target) spectra. Decoy spectra selected by the searching tool are filtered out. The number of target SSMs and decoy SSMs at a specific score can be used to compute the FDR. 
Typically, an FDR threshold of 1\% is used to minimize the number of incorrect identifications. The performance of different search tools can be compared by \textit{the number of identified spectra} at a fixed FDR threshold. 

A \textbf{standard searching} strategy can identify directly matching spectra. It assumes that precursor \mz of query and matched reference spectra are similar (narrow precursor \mz tolerance). However, as spectral libraries mainly contain unmodified reference spectra, they cannot be used to identify modified ones. Modified ones have a different intact mass, as the modifications induce mass shifts. 
\textbf{Open modification searching (OMS)} addresses these issues by (i) using a wide precursor \mz tolerance that exceeds mass shifts induced by modifications to select reference candidates~\cite{Na2015}, and (ii) using alternative spectrum similarity measures that take peak shifts due to modifications into account~\cite{bittremieux2018fast}.
Using a wide precursor \mz tolerance enables finding (partial) matches between unmodified reference spectra and their modified variants. However, a large number of candidates need to be evaluated for each query spectrum, which can be computationally demanding.

\smallsqueezeup
\subsection{Accelerated Spectral Library Searching }
OMS has recently become an increasingly popular search strategy, and there have been several studies to accelerate searches on different hardware platforms. On CPU, MSFragger~\cite{msfragger} uses a fragment ion index to retrieve matching reference spectra. 
ANN-SoLo~\cite{bittremieux2018fast, annsologpu} is a state-of-the-art OMS tool that uses GPU-powered nearest neighbor searching library, Faiss~\cite{faiss}, to prune the search space. 
Several studies have focused on accelerating spectral library searching using GPUs for efficient spectrum--spectrum similarity computation~\cite{Baumgardner2011, Li2014}. \cite{yang2019complete} used a CPU-FPGA architecture in which multiple FPGAs are used for scalability and parallelism. However, none of these studies have addressed the OMS challenges.

\begin{figure}[t]
	\centering
	\includegraphics[width=.8\linewidth]{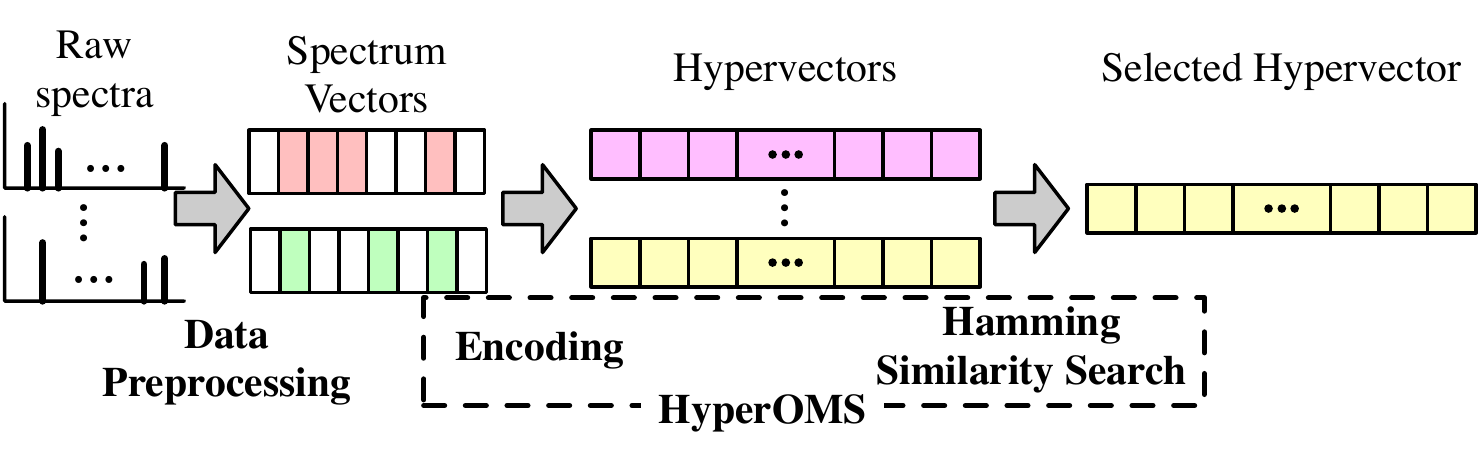}
	\caption{Overview of OMS process using \Design.} \squeezeup
	\label{fig:alg_overview}
\end{figure}

\begin{figure*}[t]
	\centering
	\begin{subfigure}[t]{0.29\linewidth}
		\includegraphics[width=\textwidth]{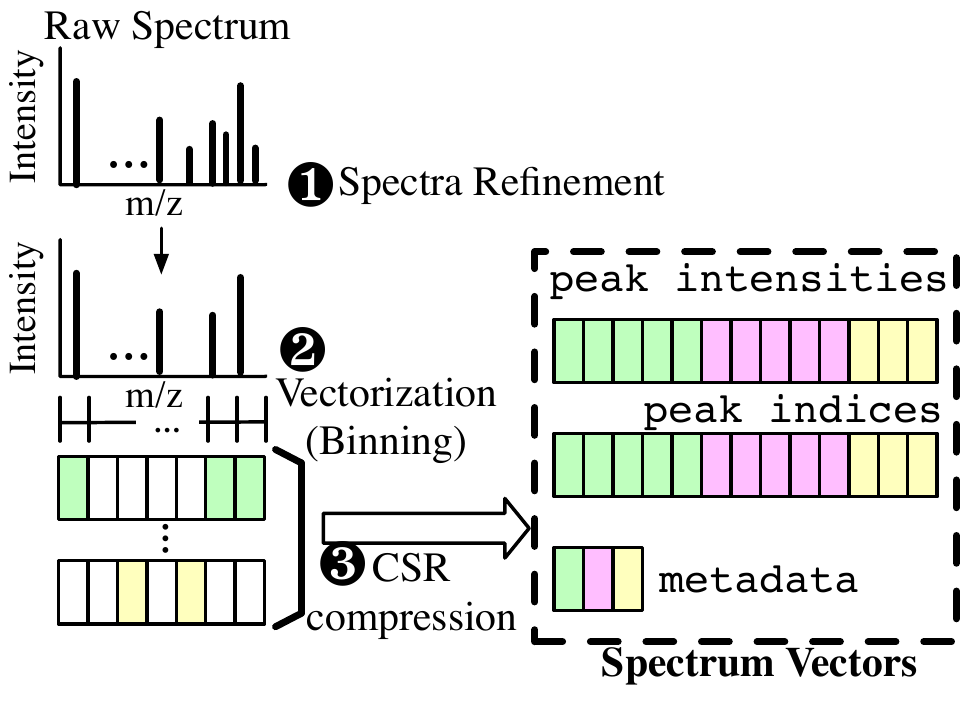} 
		\caption{Data preprocessing.}
	\end{subfigure}
	\begin{subfigure}[t]{0.21\linewidth}
		\includegraphics[width=\textwidth]{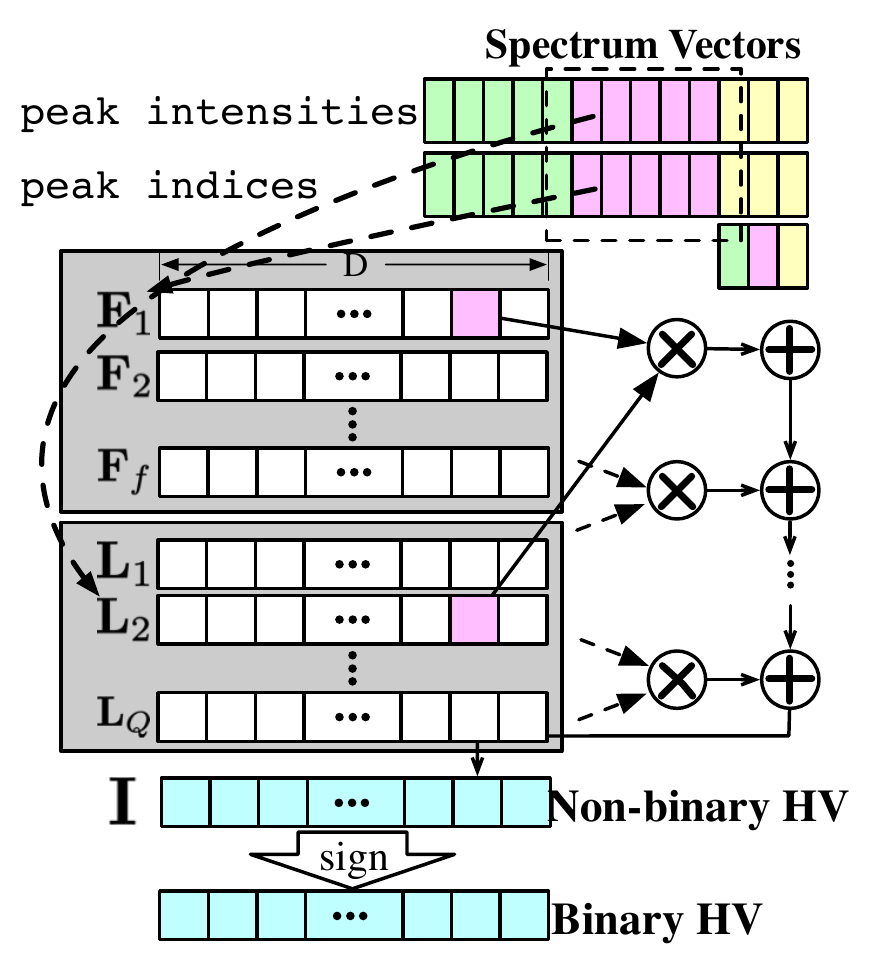} 
		\caption{Encoding.}
	\end{subfigure}
	\begin{subfigure}[t]{0.48\linewidth}
		\includegraphics[width=\textwidth]{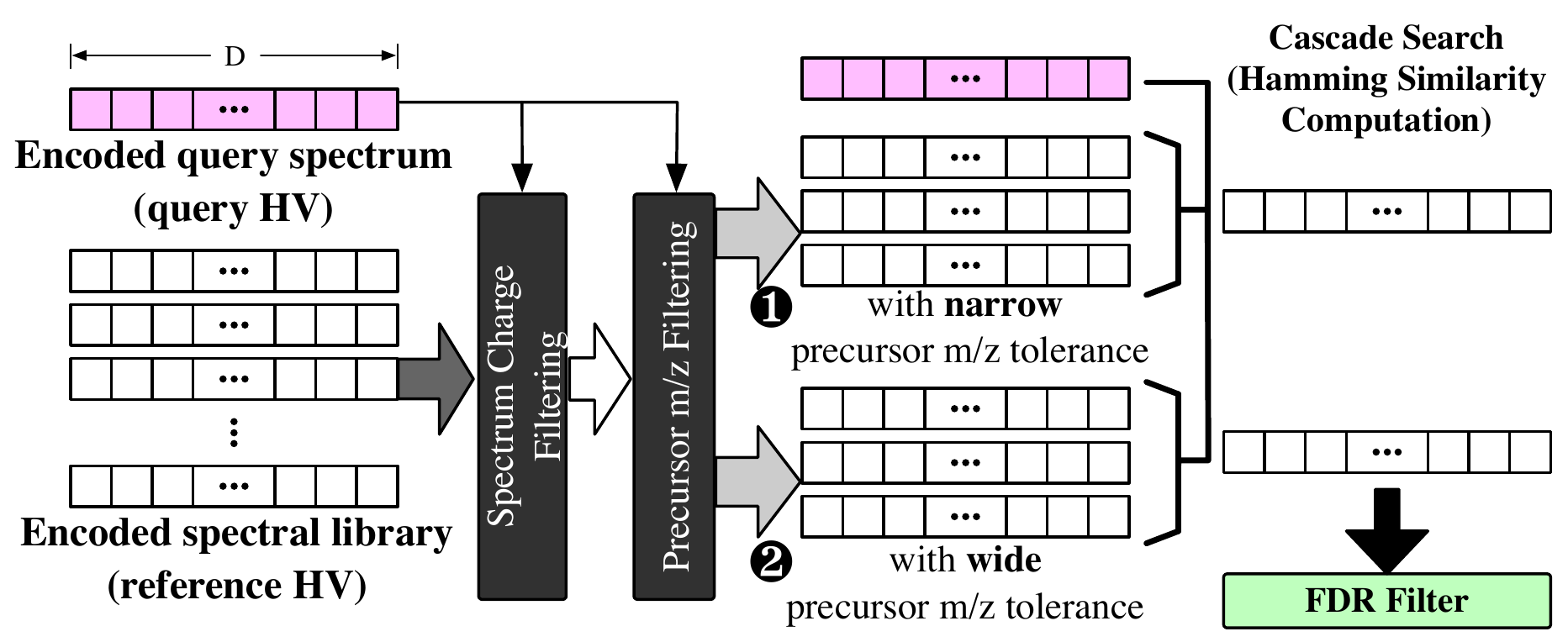} 
		\caption{Hamming similarity searching.}
	\end{subfigure}
	\caption{Data preprocessing and stages of \Design algorithm.}
	\squeezeup
	\label{fig:stages}
\end{figure*}

\smallsqueezeup
\section{\Design Algorithm}\label{sec:overview}
\Design encodes raw spectral data to a binary HD vector called \textit{hypervector} (HV) to capture the position and intensities of peaks while considering the spatial locality and value locality. Although peaks are shifted due to PTMs, the similarity between a query spectrum and a matching reference spectrum does not change dramatically. 
Furthermore, since binary vector representation is used, \Design replaces complex similarity metrics in existing OMS tools with a simple Hamming similarity operation.

\smallsqueezeup
\subsection{Overall Flow}
Figure \ref{fig:alg_overview} shows a flow of \Design. It starts with a data preprocessing step, a common step for OMS. It refines and vectorizes raw spectra and compresses them, resulting in spectrum vectors. 
\Design encodes spectrum vectors into HV during the encoding step. Next, the Hamming similarity searching step filters reference spectra according to the query's precursor \mz and spectrum charge and computes Hamming similarity between query HVs and candidate reference HVs.

\smallsqueezeup
\subsection{Data Preprocessing}
The preprocessing step (1) refines the raw spectra by removing redundant peaks and (2) vectorizes refined spectra (see Figure~\ref{fig:stages}(a)).
First, raw spectra are refined to gather meaningful peaks ($\invcircledast{1}$). We remove peaks whose intensity is below 1\% of the most intense peaks. Low-intensity peaks are considered noise. In turn, we retain 
$50$ to $150$ most intensive peaks of the spectra. Existing studies~\cite{annsologpu, spectrast} have shown that we can effectively refine spectra in this manner.

Next, we vectorize the filtered spectra ($\invcircledast{2}$). The peaks are discretized by binning the \mz range to represent a spectrum into a sparse vector of floating-point intensity values, called a \textit{spectrum vector}. 
If multiple peaks are assigned to the same \mz bin, we sum their intensity values. A large bin width can lead to loss of information when peaks are grouped into a single bin. 
For example, the mass range between 0 \mz and 2000 \mz and bin width $0.04$ (based on the resolution of the mass spectrometer) results in a dimensionality of $50,000$.
The resulting spectrum vectors have sparsity less than $1\%$. There are $50$ to $150$ peaks for each spectrum vector, and its dimensionality is $20,000$ to $50,000$. As such, we compress spectrum vectors in a compressed sparse row (CSR) format ($\invcircledast{3}$). 
The preprocessing step is normally run offline, and the resulting data is stored as a binary file for future use.
In the following, we discuss the \Design algorithm, which first encodes spectrum vector to HV and performs searching on HV.

\smallsqueezeup
\subsection{Encoding: Spectrum Vectors to Hypervectors}\label{sec:overview_encoding}

\Design encodes the data into a \textit{binary vector representation}, which can enhance the computation efficiency. There have been several efforts to represent raw data in an HD binary vector, using Locality Sensitive Hashing (LSH)~\cite{Koga2006, Norouzi2012, imani2020dual} or HDC~\cite{imani2017voicehd,kim2018efficient,salamat2019f5,imani2020dual}. 
However, these strategies do not reflect the characteristics of OMS, including peak shifts and intensity changes. For example, they treat each feature position (corresponding to peak indices in spectrum vector) as orthogonal. Peak shifts can lead to significant changes in similarity.
Conversely, our encoding takes both \textit{spatial locality} (for peak shift) and \textit{value locality} (for peak intensity change from instrument error or noise) of each feature into account. 
As a result, we can preserve similarity despite peak changes, which is essential for OMS.

\begin{figure}[t]
	\centering
	\begin{subfigure}[t]{0.48\linewidth}
		\includegraphics[width=\textwidth]{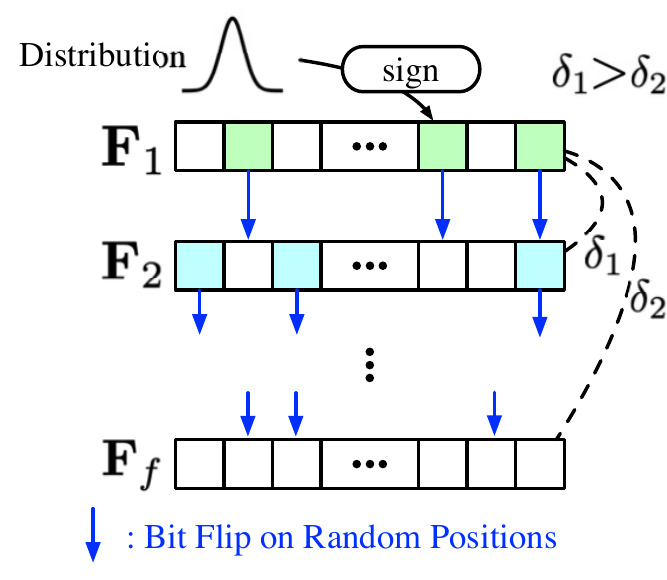}
		\caption{}
	\end{subfigure}
	\begin{subfigure}[t]{0.45\linewidth}
		\includegraphics[width=\textwidth]{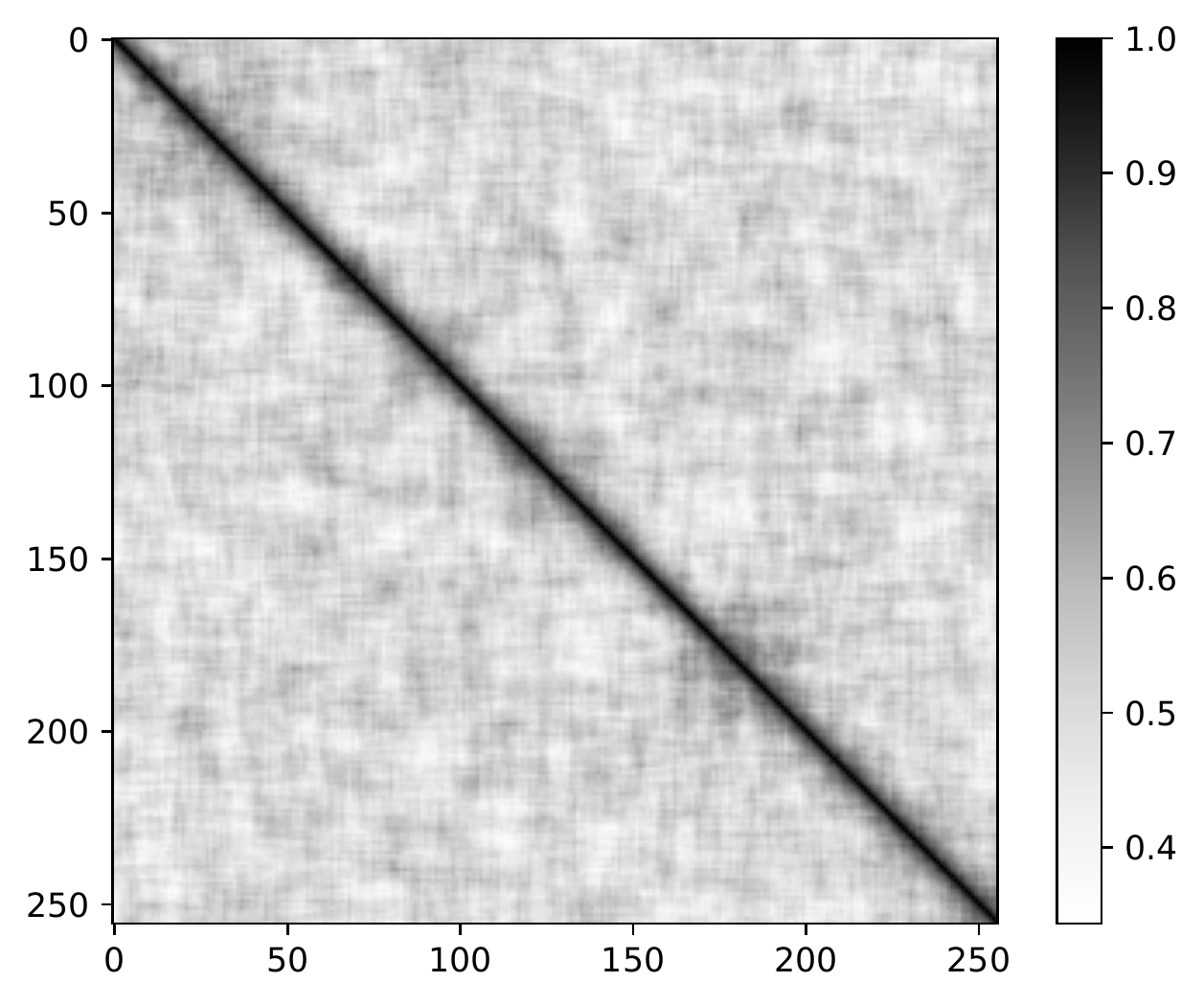}  
		\caption{}
	\end{subfigure}
	\caption{Position HV generation. (a) Strategy overview. (b) Pairwise similarity (Hamming similarity normalized by the HV dimension size) between position HVs.}
	\squeezeup
	\label{fig:bitflip_idhv}
\end{figure}

Figure~\ref{fig:stages}(b) shows the encoding process of \Design. 
Unique \textit{position HVs} $\mathbf{F}$ are assigned for each index in a spectrum vector, i.e., $\mathbf{F}_i$ corresponds to index $i$, and $\mathbf{F} \in \{\mathbf{F}_1, \mathbf{F}_2, \ldots, \mathbf{F}_f\}$. Note that $f$ is the dimensionality of spectrum vector. 
Similarly, we use \textit{level HVs} $\mathbf{L}$ to capture different intensity values in each index. We quantize intensity range into $Q$ levels, and $\mathbf{L}_i$ is assigned to each quantization level $i$ where $i \in [0, Q]$. Given two sets of HVs, $\mathbf{F}$ and $\mathbf{L}$, a spectrum vector is encoded into a HV $\mathbf{I}$ as follows. Let $\mathds{P}$ be the set of peaks in the spectrum vector, consisting of tuples $(i, j)$, with $i$ the peak index and $j$ the step value of its intensity. $\mathbf{I}$ is computed as follows:
\begin{equation}\label{eq:idlv}
	\mathbf{I} = \sum_{(i, j)\in \mathds{P}} \mathbf{F}_i \odot \mathbf{L}_j,
\end{equation}
where $\odot$ indicates element-wise multiplication.
In turn, we binarize the $\mathbf{I}$ for the computational efficiency on hardware; all positive elements are mapped to $1$ and $-1$ otherwise. 
The final representation of the HV is a binary vector. 
Spectrum vectors corresponding to the query and reference spectra are encoded to query HVs and reference HVs, respectively.
Encoding of reference spectra is done only once. The reference HVs is reused for subsequent runs since they are already identified and unlikely to change. 

\subsubsection{Reflecting Spatial Locality}
\label{subsubsec:spatial_locality}
We devise the new position HV generation method to reflect the spatial locality.
Previous studies~\cite{salamat2019f5, salamat2020accelerating} have used a permutation-based or random generation method, which makes $\mathbf{F}_i$ and $\mathbf{F}_j$ ($i \ne j$) nearly orthogonal. However, they are vulnerable to peak shifts as they accompany changes in $i$.

Figure~\ref{fig:bitflip_idhv}(a) shows the proposed position HV generation strategy. We randomly generate $\mathbf{F}_1=\{+1,-1\}^D$. In turn, we flip $\alpha$ components in random positions. 
As more flips occur, the similarity between the original vector and the flipped vector decreases. For example, the similarity ($\delta_1$) between $\mathbf{F}_1$ and $\mathbf{F}_2$ is larger than the similarity ($\delta_2$) between $\mathbf{F}_1$ and $\mathbf{F}_f$. The \Design encoding reflects the characteristics of peak shifts in OMS well: (1) neighboring positions should have spatial locality to deal with peak shifts, while (2) distant positions need to have adequate orthogonality.
The peak shift changes the index value corresponding to the intensity in the spectrum vector. Position HVs do not change significantly even if peak shifts occur; the resulting representation can be tolerable to PTMs.
As depicted in Figure~\ref{fig:bitflip_idhv}(b), for $\mathbf{F}_i$ and $\mathbf{F}_j$, the pairwise similarity has a high value when $i\approx j$ and is maximum when $i=j$ (diagonal elements). Note that we scaled down the $f$ to $128$ and $D$ to $256$ for better visibility.

\subsubsection{Reflecting Value Locality}
\label{subsubsec:value_locality}
The intensity information of the spectrum vectors is captured. We use the level HV generation method from \cite{imani2017voicehd,kim2018efficient,salamat2019f5}.
We allocate a single bit to each of the HV components, i.e., $\mathbf{L}_i \in \{-1, 1\}^D$. 
$\mathbf{L}$ that is assigned to each quantization level needs to reflect the closeness of the intensity, i.e., value locality. The similarity between $\mathbf{L}_i$ and $\mathbf{L}_{i+1}$ is higher than the similarity between $\mathbf{L}_{i}$ and $\mathbf{L}_{i+100}$. 
For instance, for the target level $p$ in percentage, $\mathbf{L}_p$, we can represent this by flipping $(D/2)\times(p/100)$ elements of $\mathbf{L}_0$.

\smallsqueezeup
\subsection{Hamming Similarity Search}
After the encoding step, \Design finds the matched reference HV that is most similar to the query HV. 
It uses \textit{Hamming similarity} (defined by the number of equal components in vector pairs) as a similarity metric.
Here, reference spectra that need to be compared primarily need to satisfy spectrum charge and precursor \mz condition per query as discussed in Section \ref{sec:background}.
Each spectrum has its spectrum charge (+2, +3, $\ldots$) and precursor \mz value. 
We gather reference spectra that (1) have the same spectrum charge as the query spectra and (2) satisfy the precursor \mz tolerance (precursor \mz difference between query and reference) condition. 

OMS assumes that precursor \mz of selected reference spectra and query spectra can have a large difference.
In other words, a wide precursor \mz tolerance is used to match modified spectra to their unmodified variants. 
However, we may miss the case of a reference spectrum with a similar precursor \mz that can pass through the FDR filter with high similarity.
To avoid such misidentifications, we adopt \textit{cascade search}~\cite{Kertesz-Farkas2015}. A narrow precursor \mz tolerance is used for standard search and FDR filtration is applied (Figure~\ref{fig:stages}(c)-$\invcircledast{1}$). In turn, remaining unidentified spectra are processed with a wide precursor \mz tolerance (Figure~\ref{fig:stages}(c)-$\invcircledast{2}$).

\smallsqueezeup
\section{Evaluation}\label{sec:evaluation}

\subsection{Methodology}
The evaluation was performed on a system equipped with Intel i7-8700K with 64GB RAM and NVIDIA Geforce GTX 1080Ti. We have implemented the \Design algorithm on GPU based on \cite{kang2022xcelhd, kang2022openhd}. To maximize the computation efficiency, we represent binary HV as a $32$bit integer array using bit packing, and similarity score computation is done by CUDA intrinsic ($\texttt{\_\_popc}$).
Since the GPU has limited memory, we split the reference and the query data into batches. We set the batch size to use the maximum amount of VRAM for GPU-based solutions. 
We measured the energy consumption of the CPU and GPU using Intel Power Gadget and \texttt{nvidia-smi}, respectively.

\noindent
\textbf{Workloads.} We evaluated \Design on two real-world datasets. Small scale dataset combines yeast spectral library~\cite{Selevsek2015} with the human HCD spectral library (total spectra: $1,162,392$) as the reference libraries, and the iPRG2012 dataset~\cite{Chalkley2014} (total spectra: $15,867$) as a query. For a larger-scale evaluation, we used the human spectral library~\cite{massivekb, wang2018assembling} (total spectra: $2,992,672$) and a HEK293 (Human Embryonic Kidney 293) dataset~\cite{Chick2015} (total spectra per query: $46,665$ on average), as reference and query spectra, respectively. Note that HEK293 consists of multiple query files (b1906$\sim$b1938).

We preprocessed query and reference spectra in a similar fashion to existing works~\cite{annsologpu,bittremieux2018fast,spectrast}, using the widely used configurations listed in Table~\ref{tab:preprocessing}. We removed the duplicates and added decoy spectra with the same ratio as the existing spectral libraries using the shuffle-and-reposition method~\cite{Lam2009} for FDR filtering.

\begin{table}[t]\small
	\centering
	\caption{Spectrum preprocessing settings.}
	\begin{tabular}{|c|c|c|}
		\hline
		& \multicolumn{2}{c|}{\textbf{Dataset}} \\ \hline
		\textbf{Parameter Name} & \textbf{Small scale} & \textbf{Large scale} \\ \hline
		\hline
		Max peaks in spectra & \multicolumn{2}{c|}{50} \\ \hline
		Min / max \mz & \multicolumn{2}{c|}{101 / 1500} \\ \hline
		Bin size & 0.05 & 0.04 \\ \hline
		Precursor \mz tolerance (narrow) & 20ppm & 5ppm \\ \hline
		Precursor \mz tolerance (wide) & 500Da & 500Da \\ \hline
	\end{tabular} 
	\squeezeup
	\label{tab:preprocessing}
\end{table}

\noindent
\textbf{Benchmarks.} 
We compare the search quality of \Design to existing search tools, including (1) SpectraST~\cite{spectrast, spectrastoms} that run on CPU and (2) the state-of-the-art OMS tool, ANN-SoLo, running on CPU~\cite{bittremieux2018fast} and GPU~\cite{annsologpu}.
We count the number of identifications to compare the search quality. All search results are evaluated at fixed $1\%$ FDR threshold, using Pyteomics~\cite{Goloborodko2013}. 

\begin{figure}[t]
	\centering
	\begin{subfigure}[t]{.7\linewidth}
		\includegraphics[width=\textwidth]{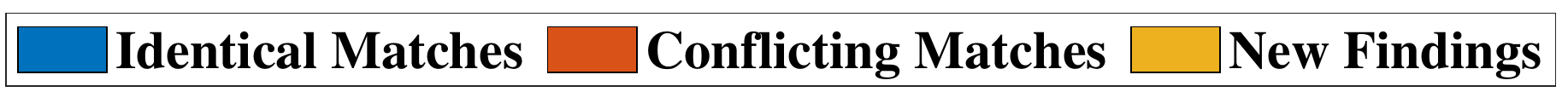}
	\end{subfigure}
	\setcounter{subfigure}{0} \\
	\vspace{-.8\baselineskip}
	\bigskip
	\begin{subfigure}[t]{0.25\linewidth}
		\includegraphics[width=\textwidth]{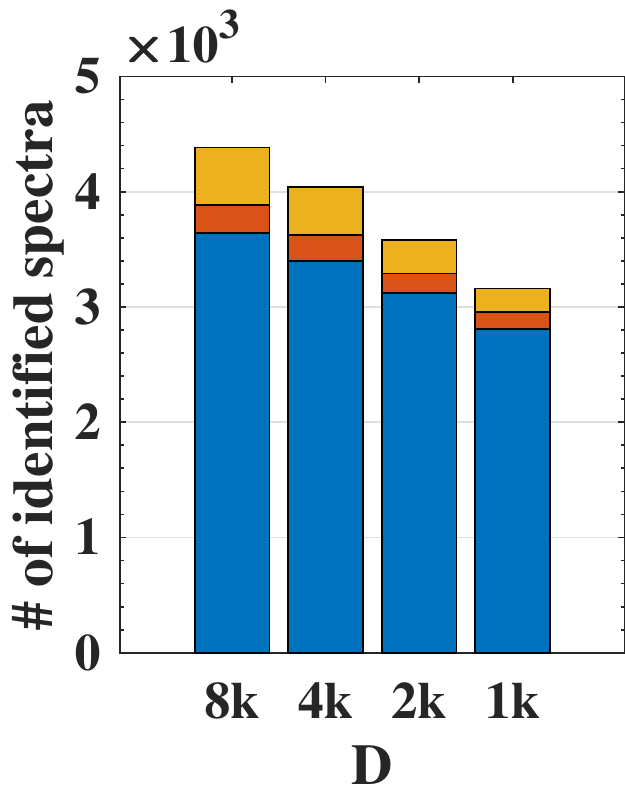}
		\caption{}
	\end{subfigure}
	\begin{subfigure}[t]{0.25\linewidth}
		\includegraphics[width=\textwidth]{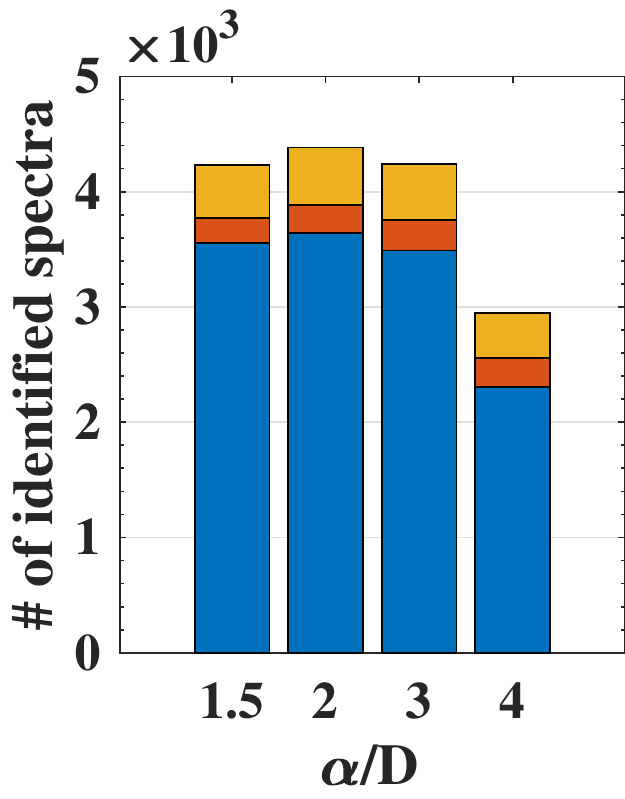} 
		\caption{}
	\end{subfigure}
	\begin{subfigure}[t]{0.22\linewidth}
		\includegraphics[width=\textwidth]{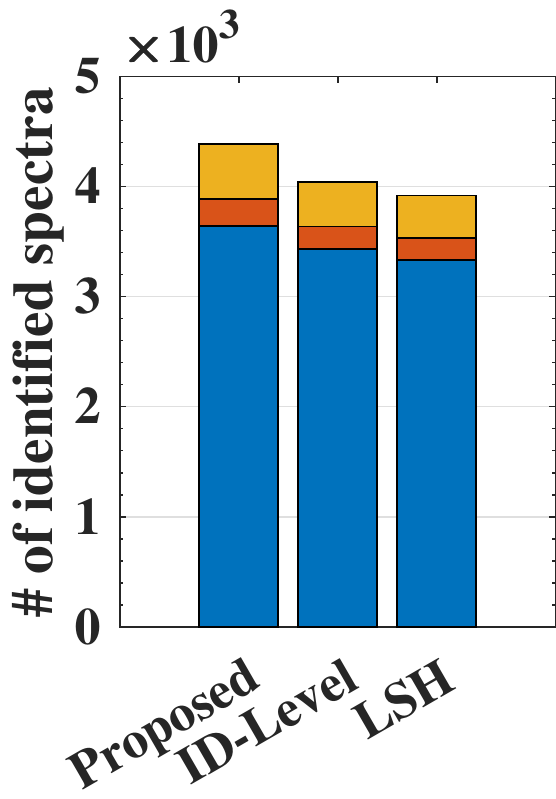} 
		\caption{}
	\end{subfigure}
	\begin{subfigure}[t]{0.2\linewidth}
		\includegraphics[width=\textwidth]{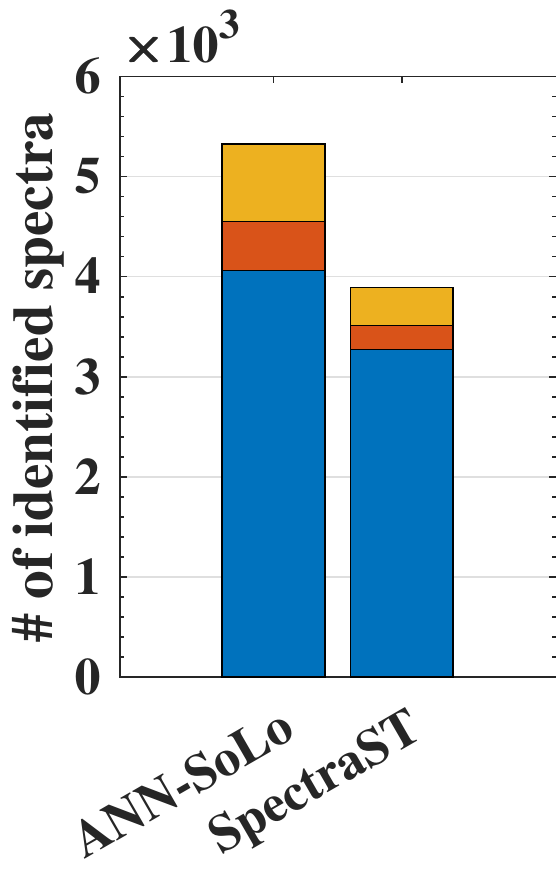} 
		\caption{}
	\end{subfigure}
	\caption{Search quality comparison on the small scale dataset. (a) \Design with different HV dimensionality. (b) \Design with different number of bit flips. (c) \Design with various encoding strategies. (d) Results of baseline tools. }
	\squeezeup
	\label{fig:consensus}
\end{figure}

\smallsqueezeup
\subsection{Impact of Encoding Configuration}\label{sec:evaluation_config}
\noindent
\textbf{HV dimensionality.}
The HV dimensionality ($D$) plays a critical role in search quality. A low $D$ limits separability. 
Figure~\ref{fig:consensus}(a) describes the impact of $D$.
The higher $D$ leads to a higher number of identifications. However, the excessive $D$ leads to an increase in computation and capacity demand. We set the $D$ to $8192$ (8k).

\noindent
\textbf{Flipped bits.}
The number of flipped bits ($\alpha$) controls the balance between orthogonality and correlation between bins. A high $\alpha$ increases the orthogonality of each position, and a low $\alpha$ helps a more number of adjacent bins to have a correlation (spatial locality). We measured the number of identifications according to the ratio of flips to $D$, i.e., $\alpha/D$. 
As shown in Figure~\ref{fig:consensus}(b), an adequate $\alpha/D$ leads to high search quality. But if we flip a small number of bits, \Design cannot clearly differentiate the peak position. Also, since the peak shifts due to PTMs are not significant, spatial locality for a limited range is required. In the rest of our evaluation, we use $\alpha=D/2$, which shows the highest performance in most cases.

\noindent
\textbf{Encoding strategy.} 
We compare the search quality of \Design with the different binary encoding strategies. As discussed in Section~\ref{sec:overview_encoding}, LSH~\cite{Koga2006, Norouzi2012, imani2020dual}, ID-Level HDC encoding~\cite{imani2017voicehd,kim2018efficient,salamat2019f5} can be used alternatively to map raw data to binary vectors. 
Figure~\ref{fig:consensus}(c) compares the search quality of \Design with the (1) proposed encoding method, (2) ID-Level HDC encoding that can capture the position of feature and its value, and (3) random projection-based LSH approach. Our encoding offers the best search quality compared to baselines. 

\noindent
\textbf{Quantization level.}
High quantization levels may not be flexible to the peak intensity changes due to noise and PTMs. Low $Q$ leads to low resolution in intensity capturing of the encoder. The quantization level $Q$ did not significantly affect the search quality unless it falls in $[8,32]$. Therefore, we use $Q=16$. 

\begin{figure}[t]
	\centering
	\begin{subfigure}[t]{.68\linewidth}
		\includegraphics[width=\textwidth]{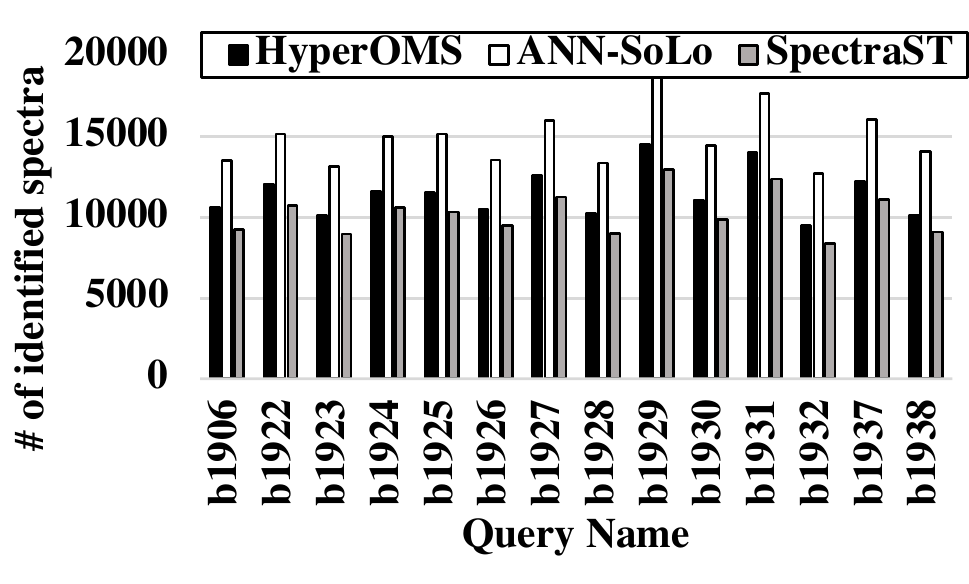} 
		\caption{}
	\end{subfigure}
	\begin{subfigure}[t]{.3\linewidth}
		\includegraphics[width=\textwidth]{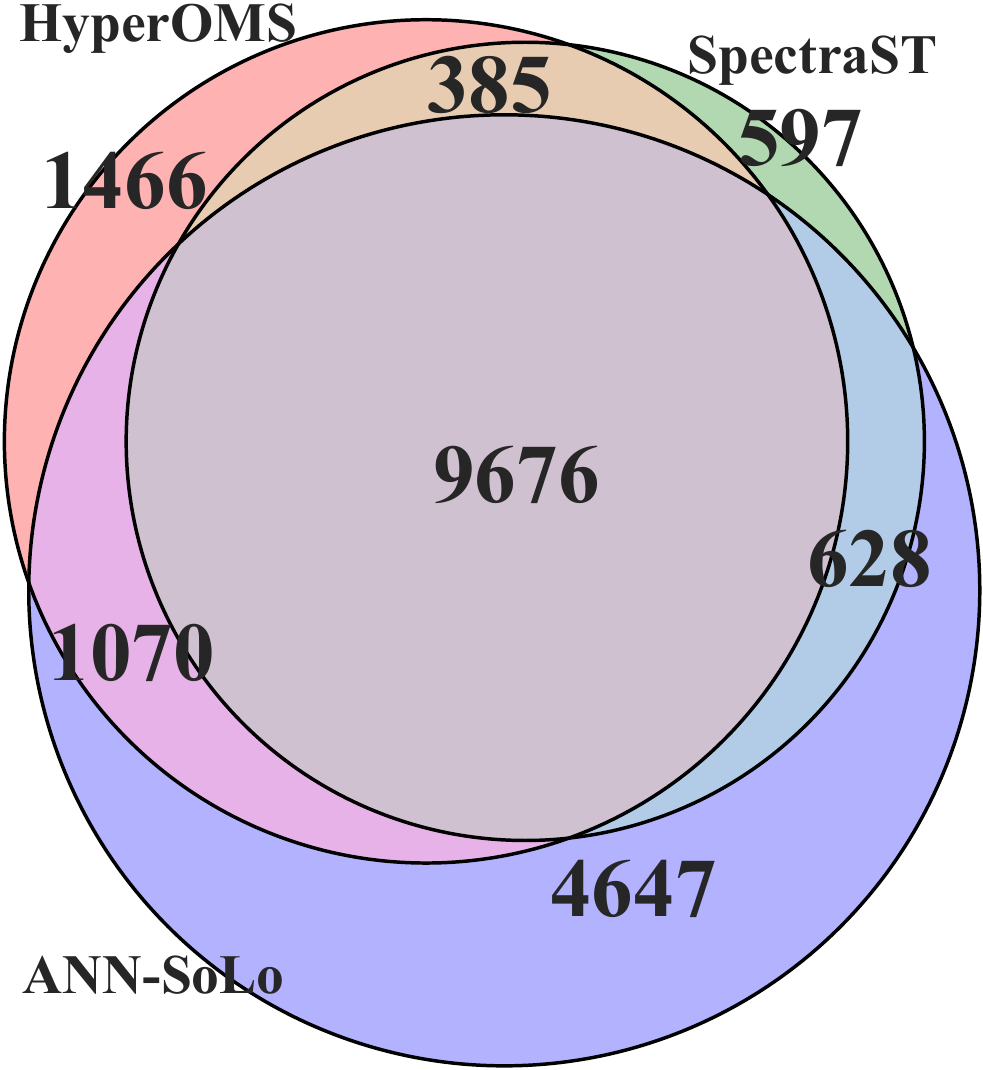}
		\caption{}
	\end{subfigure}
	\caption{Search result analysis from \Design and baseline tools on the large scale dataset. (a) Total number of identifications. (b) Venn diagram of b1927 query search result.}
	\label{fig:search_quality}
	\squeezeup
\end{figure} 

\smallsqueezeup
\subsection{Search Quality}
We searched the iPRG2012 dataset against the yeast spectral library. As no ground truth information is available when analyzing complex biological data, instead, we compare our search quality with a list of consensus identifications produced by multiple search tools during the iPRG2012 study~\cite{Chalkley2014}.
Figure~\ref{fig:consensus}(d) shows the search result of baseline tools. Among $7841$ identifications in the iPRG2012 consensus result, \Design is able to correctly identify $4141$ spectra (Figure~\ref{fig:consensus}(b)). SpectraST and ANN-SoLo manage to identify $3891$ and $5327$ identifications, respectively.

We compared the performance of \Design with the results from existing tools, including SpectraST and ANN-SoLo using the large-scale dataset. 
We used similar configurations for all tools, listed in Table~\ref{tab:preprocessing}. 
Figure~\ref{fig:search_quality} shows the number of identifications from the different search tools. 

\Design offers a higher search quality than SpectraST, i.e., \Design identifies more spectra. ANN-SoLo managed to identify more spectra than our \Design. Nevertheless, as described in Figure~\ref{fig:search_quality}(b), \Design can identify spectra that other tools can find (overlapped area). 
\Design represents spectra with an approximated form of the original data, which is robust to PTMs. Therefore, \Design can use Hamming similarity to perform OMS. Besides, ANN-SoLo uses shifted cosine similarity metric, which is accurate when finding the original spectra. 

The identification rate of \Design can be improved by increasing the HV capacity. This can be done by (1) increasing $D$ or (2) increasing the precision of each component in the HV. For example, increasing $D$ from 8k (8192) to 16k (16384) can yield up to 10\% more identifications. However, it raises the hardware cost, computational complexity, and energy consumption of the accelerator. Since our main goal is to maximize the speed and energy efficiency of the OMS while achieving reasonable quality in a biological sense, we use a binary HV with 8k dimensionality.

A ramification of lower search quality could be missing potentially relevant biomarker proteins in the context of a healthy versus diseased study, or missing data similarly impacting other downstream biological interpretations. Nevertheless, the \Design identification rate is within the range of the state-of-the-art in MS identification. For example, we can typically expect an identification rate of 33–66\% currently for human samples that we have used, and \Design satisfies the expected range criterion. 
One advantage of \Design is that there is a search quality–efficiency trade-off that can be tuned using the hyperparameters. Furthermore, the user can decide between different search engines based on their requirements. For example, \Design runs much faster with superior energy efficiency compared to existing OMS tools (Section \ref{sec:perfanalysis}). 
It could be used to efficiently process extremely large proteomics datasets consisting of tens of thousands of query files, which are being generated increasingly often recently.

\begin{figure}[t]
	\centering
	\begin{subfigure}[t]{\linewidth}
		\includegraphics[width=\textwidth]{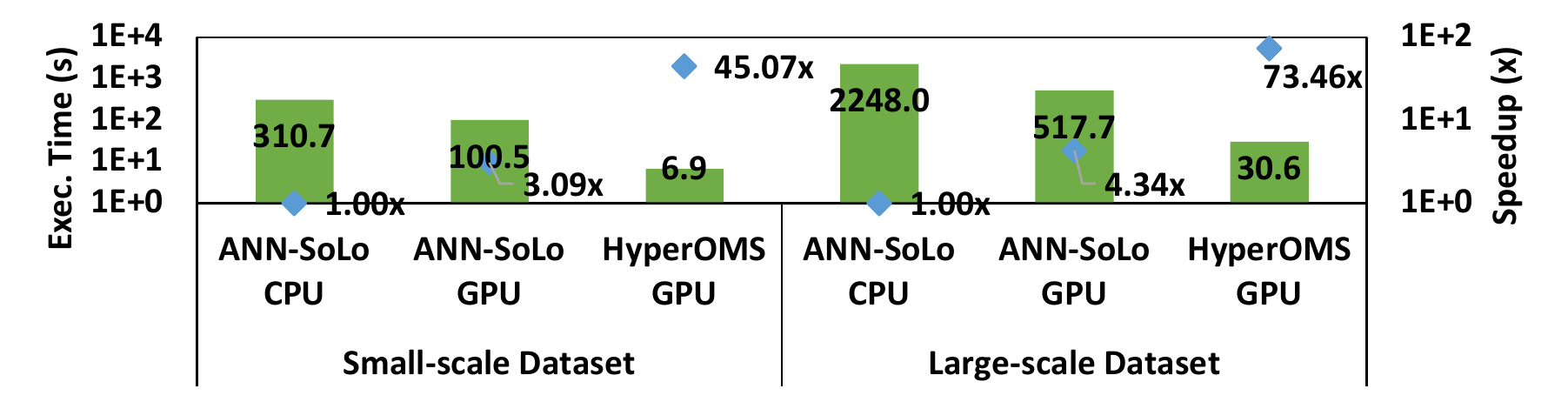} 
	\end{subfigure}\\
	\begin{subfigure}[t]{\linewidth}
		\includegraphics[width=\textwidth]{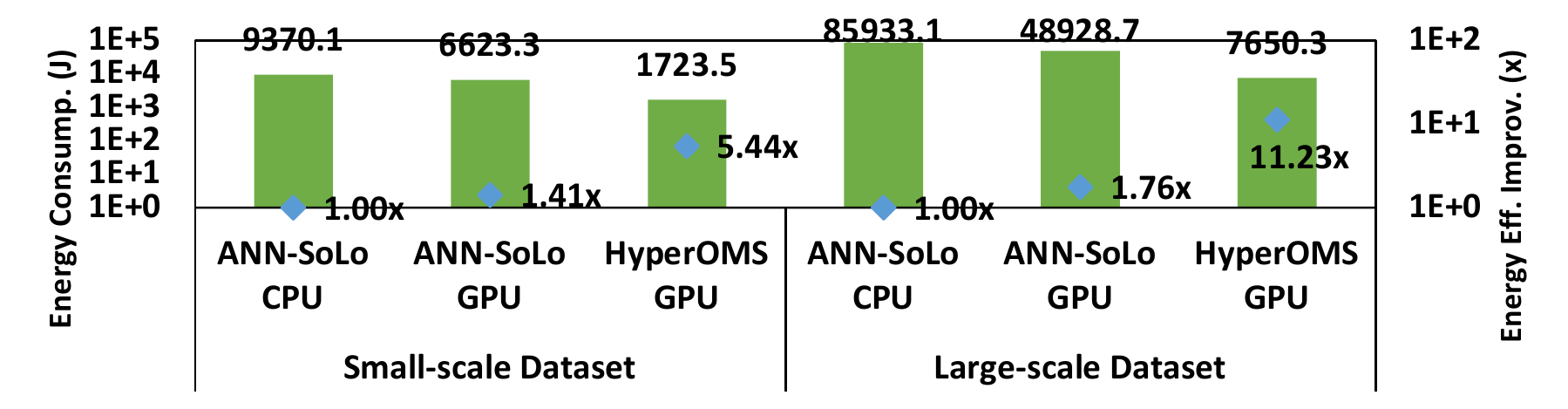}
	\end{subfigure}
	\caption{Performance and energy efficiency comparison.}
	\squeezeup
	\label{fig:perf_comparison}
\end{figure} 

\smallsqueezeup
\subsection{Speed and Energy Efficiency Improvement}\label{sec:perfanalysis}
We compare the execution time and the energy consumption of \Design on GPU to the state-of-the-art OMS tool ANN-SoLo, which offers a faster search speed with the GPU acceleration than other baseline~\cite{annsologpu}. Note that we measured the second run of the ANN-SoLo since the reference data is likely to be pre-encoded in reality. ANN-SoLo saves the pre-indexed information of the reference library on the first run and reuses it in the subsequent run. For the large-scale dataset, we averaged the measurements from multiple queries (b1906 $\sim$ b1938).

Figure~\ref{fig:perf_comparison} compares the end-to-end runtime. 
ANN-SoLo builds the index on the CPU while the encoding of \Design is done on the GPU. The \Design encoding is parallelized over HV dimensions and datapoints. 
The encoding stage of \Design, which corresponds to the index build of ANN-SoLo, is up to $8.6\times$ faster than ANN-SoLo. 
\Design uses HD binary vector and easily parallelizable Hamming similarity computation, while ANN-SoLo uses FP32 vector.
The search process of \Design GPU achieves on average $82\times$ speedup over ANN-SoLo on CPU and $11.2\times$ speedup ANN-SoLo on GPU. Ultimately, \Design GPU offers significant speedup in all stages of the OMS. 
For the end-to-end execution, \Design GPU gains an average speedup of $15.7\times$ over the state-of-the-art OMS tool running on the same GPU.

Besides, \Design running on the GPU requires more power than the ANN-SoLo, as it has high parallelism. However, the increased power consumption is compensated by reduced execution time, improving energy efficiency. 
Overall, \Design results in $7.8\times$ and $5\times$ energy efficiency improvement over CPU and GPU on average, respectively, as shown in Figure~\ref{fig:perf_comparison}.

\smallsqueezeup
\section{Conclusion}\label{sec:conclusion}
We proposed \Design, HDC-inspired massively parallel algorithm for OMS of MS-based proteomics. The proposed algorithm encodes spectra into binary HVs, considering the spatial and value locality of peaks. Therefore, \Design simplifies the execution pipeline and maximizes the computation efficiency and parallelism by using a binary vector with boolean operations. Our evaluation results show that \Design offers comparable search quality to existing OMS tools. Furthermore, \Design on NVIDIA Geforce GTX 1080Ti yields up to $17\times$ speedup and $6.4\times$ improved energy efficiency over the state-of-the-art GPU-based OMS solution.

\smallsqueezeup
\begin{acks}
	This work was supported in part by CRISP, one of six centers in JUMP (an SRC program sponsored by DARPA), SRC Global Research Collaboration (GRC) grant, and NSF grants \#1826967, \#1911095, \#2052809, \#2112665, \#2112167, and \#2100237.
\end{acks}

\smallsqueezeup
\bibliographystyle{ACM-Reference-Format}
\bibliography{main.bib}

\end{document}

%% file: ccs.tex
\ccsdesc[300]{Computing methodologies~Massively parallel algorithms}
\ccsdesc[300]{Applied computing~Computational proteomics}